\documentclass[twocolumn,aps]{revtex4} 

\def\XXint#1#2#3{{\setbox0=\hbox{$#1{#2#3}{\int}$}
     \vcenter{\hbox{$#2#3$}}\kern-.5\wd0}}

\usepackage{epsfig}
\begin{document} 
%\draft 
\title{Duality between quark-quark and quark-antiquark pairing \\ in 1+1 dimensional large $N$ models}  
\author{Michael Thies\footnote{Electronic address: thies@theorie3.physik.uni-erlangen.de}} 
\address{Institut f\"ur Theoretische Physik III\\ 
Universit\"at Erlangen-N\"urnberg\\ 
Staudtstra\ss e 7\\ 
D-91058 Erlangen\\ 
Germany} 
\date{\today} 
\begin{abstract}
We identify a canonical transformation which maps the chiral Gross-Neveu model onto a recently
proposed Cooper pair model. Baryon number and axial charge are interchanged. The same physics
can be described either as chiral symmetry breaking (quark-antiquark pairing) or as superconductivity
(quark-quark pairing).
\end{abstract} 

\pacs{11.10.Kk}
%string pacs\{\}} should always be input, 
%even if empty.} 
%\narrowtext 
\maketitle

A few years ago, Chodos, Minakata and Cooper \cite{R1} proposed a 1+1 dimensional field theoretical
model which can be solved in the $1/N$-expansion and illustrates nicely the phenomenon of Cooper pairing.
Independently, the same model has been analyzed by Kleinert and Babaev \cite{R1a} at finite $N$ in
$2+\epsilon$ dimensions, in a somewhat different context.
It is closely related to the Gross-Neveu model \cite{R2} from which it differs through
the particular form of the four-fermion coupling,
\begin{equation}
{\cal L} = \bar{\psi}^{(i)} {\rm i} \partial \!\!\!/ \,\psi^{(i)} +
2 G^2 (\bar{\psi}^{(i)} \gamma_5 \psi^{(j)})  (\bar{\psi}^{(i)} \gamma_5 \psi^{(j)}) \ .
\label{eq1p}
\end{equation}
Due to the arrangement of flavour indices, the vacuum develops a $\langle \psi \psi \rangle$
condensate and breaks spontaneously the $U(1)_V$ symmetry (baryon number),
at least in the large $N$ limit. 
Other properties
at zero temperature and chemical potential derived in Ref. \cite{R1} are strongly reminiscent of the
Gross-Neveu model, 
namely asymptotic freedom, dynamical fermion mass generation and a massless ``would be" Goldstone boson
(here: a ``diquark").
Later on, this model has been augmented by a scalar-scalar interaction of  
Gross-Neveu type \cite{R4}. The competition between $\langle \bar{\psi} \psi \rangle$
and $\langle \psi \psi \rangle$ condensates was studied in detail at finite temperature and
chemical potential. Given the current interest in color superconductivity in
the context of Quantum Chromodynamics (for a recent review, see e.g. \cite{R5}), such a
pedagogical model is highly welcome.

In this short note, we would like to clarify the relation between the Cooper pair model \cite{R1}
(hereafter referred to as ``BCS$_2$ model") and the standard Gross-Neveu model with continuous chiral symmetry
(``NJL$_2$ model") \cite{R2,R6}. Confirming the results of Ref. \cite{R1a}, we will show that at zero
chemical potential the two theories are equivalent; they can in fact be mapped into each other
by a canonical transformation. At non-zero chemical potential the two
theories are no longer equivalent but nevertheless closely related. This also sheds
some new light onto a recent work by Ohwa \cite{R7}. This author finds a chiral crystal structure of baryonic matter
in the NJL$_2$ model, confirming Ref. \cite{R8}, whereas he sees no violation of translational invariance
in the BCS$_2$ model. 

We choose the following representation for the $\gamma$-matrices,
\begin{equation}
\gamma^0 = \sigma_1 \ , \qquad \gamma^1 = - {\rm i} \sigma_2 \ , \qquad \gamma_5 = \sigma_3 \ .
\label{eq2p}
\end{equation}
In this representation, the upper and lower components of a Dirac spinor describe right-handed
and left-handed fermions. The interaction Lagrangian of the BCS$_2$ model, Eq. (\ref{eq1p}),
can then be rearranged into the form
\begin{equation}
{\cal L}_{\rm int} = 4 G^2 \psi_R^{(i)\dagger} \psi_L^{(i)\dagger} \psi_L^{(j)} \psi_R^{(j)} \ .
\label{eq3p}
\end{equation}
Let us compare this to the original NJL$_2$ model \cite{R2} with the familiar Lagrangian
\begin{equation}
{\cal L}= \bar{\psi}^{(i)} {\rm i} \partial \!\!\!/ \, \psi^{(i)} +
\frac{g^2}{2}\left[ (\bar{\psi}^{(i)}\psi^{(i)})^2- (\bar{\psi}^{(i)} \gamma_5 \psi^{(i)})^2 \right] \ .
\label{eq4p}
\end{equation}
In this case, the interaction part can be rewritten as 
\begin{equation}
{\cal L}_{\rm int}' = 2 g^2 \psi_R^{(i)\dagger} \psi_L^{(i)} \psi_L^{(j)\dagger} \psi_R^{(j)} \ .
\label{eq5p}
\end{equation}
The two interaction Lagrangians (\ref{eq3p}) and (\ref{eq5p}) are mapped onto
each other by the transformation
\begin{equation}
\psi_L^{(i)} \leftrightarrow \psi_L^{(i)\dagger} \ ,
\label{eq6p}
\end{equation}
provided we identify $g^2$ with $2G^2$.
The free part of the Lagrangian in the above representation reads
\begin{equation}
{\cal L}_0 = \psi_R^{(i)\dagger} {\rm i}\left( \partial_0 + \partial_1\right) \psi_R^{(i)} +
 \psi_L^{(i)\dagger} {\rm i}\left( \partial_0 - \partial_1\right) \psi_L^{(i)} \ . 
\label{eq7p}
\end{equation}
The free action is invariant under the transformation (\ref{eq6p}), since a minus sign
from the different ordering of the Grassmann variables is compensated by a minus
sign from partial integration. In view of thermodynamical applications, we should also
include a chemical potential. In general, one can choose independent chemical
potentials for right-handed and left-handed fermions or, equivalently, for U(1)$_V$ and
U(1)$_A$ charges. This correponds to adding
\begin{equation}
\delta {\cal L} = \mu \left( \psi_R^{(i)\dagger} \psi_R^{(i)} +\psi_L^{(i)\dagger} \psi_L^{(i)}\right)
+ \mu_5  \left( \psi_R^{(i)\dagger} \psi_R^{(i)} - \psi_L^{(i)\dagger} \psi_L^{(i)}\right)
\label{eq8p}
\end{equation}
to the Lagrangian. Applying the transformation (\ref{eq6p}) to $\delta {\cal L}$
is then tantamount to interchanging $\mu$ and $\mu_5$. 

Since the transformation (\ref{eq6p}) can be realized by a change of integration
variables in the path integral, we have thus found a correspondence between the BCS$_2$ model
with chemical potentials $(\mu, \mu_5)$ and the NJL$_2$ model with chemical potentials $(\mu_5,\mu)$.
This is a kind of ``duality" between a theory with fermion-fermion pairing and one with
fermion-antifermion pairing. The physical meaning of the transformation (\ref{eq6p}) can perhaps
more easily be understood in the canonical language. The usual expansion of the fermion field operator
in terms of annihilation operators of right-handed and left-handed fermions,
\begin{equation}
\psi^{(i)}(x) = \int {\rm d}p \left( \begin{array}{c} a^{(i)}_p \\ b^{(i)}_p \end{array} \right) {\rm e}^{{\rm i}px}\ , 
\label{eq9p}
\end{equation}
turns the mapping (\ref{eq6p}) into
\begin{equation}
a^{(i)}_p \leftrightarrow a^{(i)}_p \ , \qquad b^{(i)}_p \leftrightarrow b^{(i)\dagger}_{-p} \ ,
\label{eq10p}
\end{equation}
i.e., particle-hole conjugation for left-handed fermions only. Particle-hole conjugation as a canonical
transformation is routinely used in non-relativistic many-fermion theory. In the Hamiltonian approach to 
relativistic fermion systems, it is fundamental for the concept of a Dirac sea.  
This gives us some confidence that our transformation is indeed a legitimate one. 

After this preparation we return to Refs. \cite{R1} and \cite{R7}. A major part of Ref. \cite{R1} is
devoted to the BCS$_2$ model in the large $N$ limit at zero temperature and chemical potential. According
to the above reasoning and Ref. \cite{R1}, this model is equivalent to the NJL$_2$ model. This explains at once the findings
about the $\beta$-function (keeping in mind the redefinition of the coupling constant, $g^2\equiv
2 G^2$), the dynamical fermion mass and the massless ``diquark" (dual to the ``pion").
  
More interestingly, we
can now appeal to other known properties of the NJL$_2$ model (see e.g. the review \cite{R9}) to predict
features of the BCS$_2$ model which have not yet been addressed in Ref. \cite{R1}.
Apart from the massless ``pion", the NJL$_2$ model possesses a marginally bound
``sigma" meson of mass $2m$, where $m$ is the dynamical fermion mass. This object must also
show up in the ``diquark" spectrum of the BCS$_2$ model. Furthermore, there are massless baryons
in the NJL$_2$ model which can be described as topologically non-trivial excitations of the pion field,
in analogy to the Skyrme picture in 3+1 dimensions \cite{R9}. Baryon number can be shown to be 
the same as U(1)$_A$ winding number.
Under left-handed particle-hole conjugation (\ref{eq6p}), baryon number
goes over into axial charge.
In the dual picture of the BCS$_2$ model we therefore predict that the winding number of the U(1)$_V$ phase
is equal to the axial charge. The massless baryons of the NJL$_2$ model should manifest
themselves in the BCS$_2$ model as massless states of definite axial charge.

Let us now briefly discuss thermodynamics.
At zero chemical potential but finite temperature, we expect both theories to behave identically,
the broken U(1)$_A$ (NJL$_2$ model) and U(1)$_V$ (BCS$_2$ model) symmetries
getting restored at a common critical temperature $T_c=m{\rm e}^{\gamma}/\pi$ (see \cite{R9} and references
therein). At finite chemical potential, investigations of the BCS$_2$ model were only carried
out for $\mu_5=0$ so far. According to the duality arguments, studying the BCS$_2$ model at chemical potential
($\mu,0$) is the same as studying the NJL$_2$ model at chemical potential ($0,\mu$), i.e., zero baryon
chemical potential but non-zero ``axial chemical potential" $\mu_5=\mu$. In this sense, Ohwa's work \cite{R7}
can be re-interpreted as study of a single model (the NJL$_2$ model), but with two different chemical potentials
($\mu$ and $\mu_5$). This may be helpful in interpreting his results in more familiar terms.

Turning to generalizations of the BCS$_2$ model of the type considered in Ref. \cite{R4}, 
it is clear that the duality transformation is not of immediate help here. It maps these models onto 
other models which have also not been studied yet. From the point of view of duality, each solved model
of this class of field theories
yields the solution of another model (possibly with ``dual" chemical potentials). Thus for instance the
known large $N$ solution of the Gross-Neveu model with discrete chiral symmetry \cite{R2,R10},
\begin{equation}
{\cal L}= \bar{\psi}^{(i)} {\rm i} \partial \!\!\!/ \, \psi^{(i)} +
\frac{g^2}{2} (\bar{\psi}^{(i)}\psi^{(i)})^2 \ ,
\label{eq11p}
\end{equation} 
can be mapped onto the solution of a (so far unexplored) dual model with Lagrangian
\begin{equation}
{\cal L}'= \bar{\psi}^{(i)} {\rm i} \partial \!\!\!/ \, \psi^{(i)} +
\frac{g^2}{2} \left(\psi_R^{(i)\dagger}\psi_L^{(i)\dagger} +  \psi_L^{(i)}\psi_R^{(i)}\right)^2 \ .
\label{eq12p}
\end{equation}

Along similar lines, one can now easily write down ``self-dual" models by adding the ``dual" of
some interaction term with the same coupling constant. Such models should have
a very high degree of symmetry, degenerate $\bar{q}q$ and $qq$ states and 
perhaps other unusual properties which might be worth studying.

Summarizing, there are apparently circumstances under which Cooper pairing and
chiral symmetry breaking (or quark-quark and  quark-antiquark pairing)
are one and the same thing ---  it is a matter of convention which language one chooses
to describe a single physical phenomenon.  
In order to show this duality, we had to resort to a number of drastic simplifications 
as compared to the real world. Massless quarks, simple four-fermion interactions, 
1+1 dimensions and the large $N$ limit
were all instrumental. It would be interesting to find out whether any aspects
of this ``duality" nevertheless survive under more realistic conditions.


\newpage

\begin{references} 
\bibitem{R1}
A. Chodos, H. Minakata and F. Cooper, Phys. Lett. {\bf B449}, 260 (1999).
\bibitem{R1a}
H. Kleinert and E. Babaev, Phys. Lett. {\bf B438}, 311 (1998).
\bibitem{R2}
D.J. Gross and A. Neveu, Phys. Rev. D {\bf 10}, 3235 (1974).
\bibitem{R4}
A. Chodos, F. Cooper, W. Mao, H. Minakata, and A. Singh, Phys. Rev. D {\bf 61}, 045011 (2000).
\bibitem{R5}
K. Rajagopal and F. Wilczek, in: {\em At the frontier of particle physics: Handbook of QCD}, 
Boris Ioffe Festschrift, ed. by M. Shifman, Vol. 3, ch. 35, p. 2061, World Scientific,
Singapore (2001).
\bibitem{R6}
Y. Nambu and G. Jona-Lasinio, Phys. Rev. {\bf 122}, 345 (1961).
\bibitem{R7}
K. Ohwa, Phys. Rev. D {\bf 65}, 085040 (2002).
\bibitem{R8}
V. Sch\"on and M. Thies, Phys. Rev. D {\bf 62}, 096002 (2000).
\bibitem{R9}
V. Sch\"on and M. Thies, in: {\em At the frontier of particle physics: Handbook of QCD}, 
Boris Ioffe Festschrift, ed. by M. Shifman, Vol. 3, ch. 33, p. 1945, World Scientific,
Singapore (2001).
\bibitem{R10}
M. Thies and K. Urlichs, hep-th/0302092.
\end{references}
\end{document}